\documentclass[lettersize, journal, doublecolumn]{IEEEtran}

\usepackage[english]{babel}

\usepackage{subcaption}
\DeclareCaptionLabelSeparator{periodspace}{.\quad}
\usepackage[]{footmisc}
\usepackage{amsmath,amsfonts}
\usepackage{algorithmic}
\usepackage{algorithm}
\usepackage{array}
\usepackage{textcomp}
\usepackage{stfloats}
\usepackage{url}
\usepackage{verbatim}
\usepackage{graphicx}
\usepackage{cite}
\usepackage{multirow}
\usepackage{psfrag}
\usepackage[table,xcdraw]{xcolor}
\usepackage[normalem]{ulem}

\bgroup

\newcommand{\AK}[1]{\textcolor{black}{#1}}
\newcommand{\AKK}[1]{\textcolor{black}{#1}}

\newcommand{\MS}[1]{\textcolor{black}{#1}}

 \usepackage{tabularray}
 \UseTblrLibrary{diagbox}
 
\def\BibTeX{{\rm B\kern-.05em{\sc i\kern-.025em b}\kern-.08em
    T\kern-.1667em\lower.7ex\hbox{E}\kern-.125emX}}

\makeatletter
\newcommand{\linebreakand}{
  \end{@IEEEauthorhalign}
  \hfill\mbox{}\par
  \mbox{}\hfill\begin{@IEEEauthorhalign}
}
\makeatother

\title{V2V Path Loss Modeling at 26 GHz Based on Real-Traffic Measurements}

\author{Pawe\l~Kryszkiewicz, Adrian~Kliks,~\IEEEmembership{Senior Member,~IEEE}, Pawe\l~Sroka, Micha\l~Sybis,~\IEEEmembership{Member,~IEEE}
\thanks{The work has been realized within the project no. 2018/29/B/ST7/01241 funded by the National Science Centre in Poland. The authors are with the Institute of Radiocommunications, Poznan University of Technology, Poznan, Poland, A. Kliks is also with Luleå University of Technology, Luleå, Sweden, email: \{pawel.kryszkiewicz,adrian.kliks,pawel.sroka,michal.sybis\}@put.poznan.pl. P. Kryszkiewicz is the corresponding author. 
 For the purpose of Open Access, the author has applied a CC-BY public copyright license to any Author Accepted Manuscript (AAM) version arising from this submission.}
}

\IEEEoverridecommandlockouts

\usepackage{tikz}
\usepackage{textcomp}
\usepackage{hyperref}
\usepackage{lipsum}

\newcommand\copyrighttext{%
  \footnotesize \textcopyright 
  2024 IEEE. Personal use of this material is permitted. Permission from IEEE must be obtained for all other uses, in any current or future media, including reprinting/republishing this material for advertising or promotional purposes, creating new collective works, for resale or redistribution to servers or lists, or reuse of any copyrighted component of this work in other works. Published version: 10.1109/LWC.2024.3484312
  }
\newcommand\copyrightnotice{%
\begin{tikzpicture}[remember picture,overlay]
\node[anchor=south,yshift=5pt] at (current page.south) {\fbox{\parbox{\dimexpr\textwidth-\fboxsep-\fboxrule\relax}{\copyrighttext}}};
\end{tikzpicture}%
}
\begin{document}

\maketitle
\copyrightnotice

\begin{abstract}
In this letter, we investigate single-slope path loss models complemented with shadowing effects in the context of vehicular communications. We present several models obtained based on extensive measurement campaigns with inter-vehicle transmission conducted at 26.555 GHz in real-\AKK{traffic} experiments, mainly along high-speed roads. Particular attention has been put on the impact of aerial characteristics (omnidirectional versus directional), surrounding environment (e.g., urban versus rural), and their mounting point on cars (at the rooftop, on the bumper, and below the car chassis).
% Detailed models have been proposed and compared.in various environments for two former mounting locations.
Finally, the effect of signal ducting and of the number of blocking cars has been analyzed and the decorrelation time has been discussed.
\end{abstract}

\section{Introduction}
\label{sec:intro}
\IEEEPARstart{R}{eliable} wireless \AK{vehicle-to-everything (V2X)} communications is the basis for secure, intelligent transportation systems of the future \cite{Zheng2016, Ucar2018}. 
%On one hand side, it is evident that the better the driving performance, the less impact the transportation system has on the natural environment. On the other, however, besides efficiency, driving safety is the most important for every street user. It plays a particular role in the context of road freight forwarding, where fully- or semi-autonomous driving is envisaged nowadays. One of the straightforward applications of this idea is the so-called autonomous platooning, as discussed in, e.g., \cite{Zheng2016, Ucar2018}. 
%In such an approach, the string of cars led by a leader drives in a coordinated manner - such a group of cars is often referred to as a platoon. To reduce fuel consumption and increase road capacity, the inter-car distance should be kept sufficiently small~\cite{SARTRE}. 
%USUNIETE To achieve this goal, preventing intra-platoon cars from crashes, a reliable wireless communications have to be guaranteed, allowing for fast and efficient information exchange between the platoon members and enabling advanced cruise control. 
%USUNIETE 4 ITERACJA
%There are two main approaches to providing vehicle-to-everything (V2X) wireless data transmission in intelligent transportation systems (ITS), mainly: dedicated short-range communications (DSRC) or cellular vehicle-to-everything (C-V2X) systems. As the former one bases on the IEEE 802.11p standard \cite{IEEE80211}, in the second case, the Long Term Evolution (LTE) or New Radio (NR) standards are applied \cite{VUKAD2018}.
However, various tests have shown that existing communication technologies are sensitive to the wireless channel congestion problem \cite{VUKAD2018}.% Thus, the selection of alternative frequency bands may be necessary. One of the promising solutions is to shift the intra-platoon transmission to other frequency ranges, which tend to be less occupied, such as the millimeter wave band \cite{Sroka2020a,Naik2019}.
 However, the application of the new frequency bands for V2V \AKK{purposes} entails the need for a precise understanding of signal propagation in typical, everyday scenarios. %USUNIETE 4iteracja  In our prior works, we have discussed various aspects of this concept; for example, in \cite{Sroka2021}, the dynamic power and frequency allocation scheme for autonomous platooning has been investigated.  %Moreover, the application of radio environment maps for the same purpose has been proposed in \cite{Hoffmann2022}. Both 
\AKK{\cite{Sroka2021}} considers the TV white spaces for data transmission. Complementing it, in~\cite{Kryszkiewicz2021}, we have analyzed the impact of blocking cars on the wireless channel path loss (PL) in the frequency band around 26 GHz. 

To model the V2X system operating in sub-30 GHz band correctly and evaluate its performance reliably, the impact of the communication channel on signal propagation has to be precisely analyzed in a dynamic scenario. Various approaches are feasible here for channel modeling, such as those based on ray-tracing/ray-launching, statistical modeling, or utilizing artificial intelligence engine \cite{Hemadeh2018,Wang2018,Zheng2015}. While many models can be found for sub-6 GHz bands, there is a limited number of solutions available for higher-frequency scenarios in the V2X context, e.g. in \cite{Huang2020}. \MS{Moreover, global standardized models could also be mentioned, such as \cite{m2412} or \cite{3gpp38901}, however, these consider the cellular-specific scenario setup, thus cannot be applied directly for V2V scheme. The 3GPP TR 37.885 model \cite{3gpp37885} accounts for the specifics of V2V communications in frequencies above 6~GHz, such as low antenna height or blockage by other vehicles. However, the split between urban and highway scenarios is available only for the line-of-sight (LOS) case. Moreover, the surrounding environments (such as acoustic shields) are not considered.} 
% re is no surrounding-specific model, that accounts e.g., for the presence of acoustic screens on the highway, which, according to the outcomes of our results, has a significant impact on the propagation.}\\
%\PK{Troche brakuje tu dyskusji innych modeli mmW dla V2V. wydaje mi sie ze w softcom bylo odniesienie do raytracingu w tym przypadku. } \PS{Zgadzam się ogólnie, ale to jest letters, więc nie ma na to miejsca. Są ref-y do artykułów o tym i to moim zdaniem wystarczy}\\

 The novelty of the paper is the following:
 \begin{itemize}
 \item First, this paper further extends the investigation on the sub-30 GHz channel, focusing on its measurements and modeling for V2V communications in the realtraffic case, including high-speed mode (typically between 50 to 110 kmph) for various configurations. Thus, as a novelty, we present the PL and shadowing models for a real-traffic case, for the 26.555 GHz band using a configuration with communicating antennas located at three different mounting points: on the rooftop, on the cars' bumpers, and below the car. \item Moreover, we conduct the measurements considering: two aerial types - horn and biconical, two environment types - rural and urban, and four surrounding types (in the rural case) -  \textit{housing},\textit{ screens}, \textit{fields}, and \textit{forests}.
 \item We considered three special cases for path loss modeling: the effect of "signal ducting" below the track, blocking cars' impact, and the decorrelation time calculation. 
 \item Finally, let us emphasize that, most importantly, the measurements were performed in a real traffic scenario while driving on a highway near Poznan, Poland. This makes the obtained empirical model highly valuable for practical V2V communications applications. 
 \end{itemize}
 
%The letter is organized as follows. First, we describe the experimentation setup in Sec.~\ref{sec:expsetup}, while in Sec.~\ref{sec:plmodels}, the obtained results are presented. The paper is then concluded in Sec. \ref{sec:conclusions}. 

\section{Measurement Setup}
\label{sec:expsetup}
\begin{figure}[!t]
\centering
\includegraphics[width=3.1in]{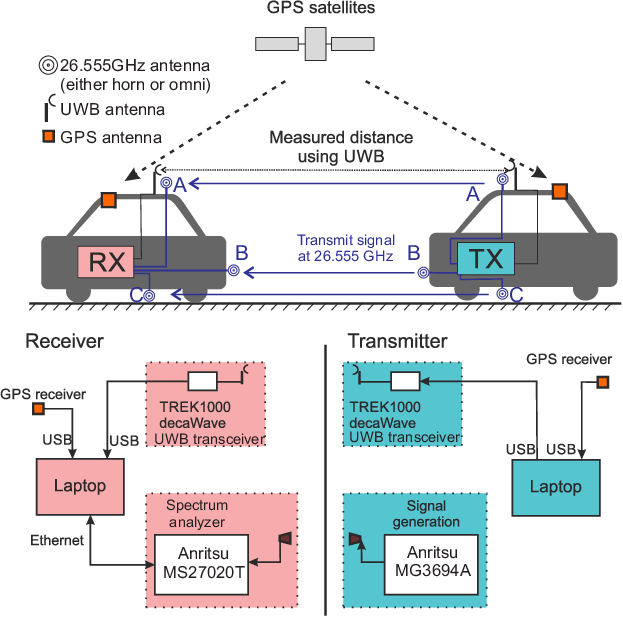}
\caption{Measurement setup used during the experiment}
\label{fig_carsetup}
\end{figure}

\IEEEpubidadjcol
To conduct the measurements, an extended version of the system setup proposed in \cite{Kryszkiewicz2021} has been considered. We have used two cars to carry the needed equipment, as presented in Fig. ~\ref{fig_carsetup}: Citroen C4 Spacetourer being the transmitting vehicle and Peugeot 5008 carrying the receiver. The transmitter has been composed of the Anritsu MG3694A signal generator power-supplied from a portable battery via the power inverter. \AK{ It has generated a single-carrier signal at the center frequency of 26.555 GHz (a sub-millimeter band, channel n257 of 5G FR2), with the transmit power set to 7 dBm. The chosen transmit power value resulted from the hardware setup's limitations, directional gains of the used antennas, and the radio transmission permit obtained from the Regulator, as we operate in a licensed band.
While we used a continuous wave for transmission, the reception bandwidth of 10 kHz used in the spectrum analyzer resulted in a low noise level. This allowed us to measure path loss higher than 130 dB with a distance of up to 1 km, i.e., higher transmit power was not needed.}

 %\PS{Zastanawiam się czy nie warto dodać, że ten zakres jest rozważany na całym świecie jako jedno z podstawowych pasm FR2 do użycia w 5G i zwykle używany do testów FR2.}
Signal reception at the receiver has been performed using an Anritsu MS2720T portable spectrum analyzer configured to a Zero Span mode with a Resolution Bandwidth filter of 10 kHz. A single span comprised 551 consecutive power samples obtained over $55 ms$. To remove fast fading effects, these samples have been averaged in linear scale \MS{to obtain a single, large-scale fading sample}. Such a measurement has been executed with a maximum speed of the utilized spectrum analyzer, i.e., approximately every $0.73~s$. Such an approach causes the mean value of averaging distance to be within the order suggested by Lee \cite{lee_condition}, that is, $81.4\lambda$ or $10.1\lambda$. %, with the shadowing effect potentially being over/underestimated.

% \MS{Moreover, to verify the W.C.Y Lee condition \cite{lee_condition}, we compared an instantaneous relative speed with the duration of the time needed to collect 551 samples ($55~ms$). Regardless of the speed used (either relative speed between transmitter and receiver or maximum value of TX speed, RX speed, and TX-RX relative speed), the mean value of averaging distance is within the order suggested by Lee, that is, $81.4\lambda$ or $10.1\lambda$.}

Both the transmitter and the receiver have been connected to antennas using low-loss coaxial cables Jyebao K30K30-53LD-40G150, applying 2.92 mm connectors. Two types of antennas have been considered in this experiment:
\begin{enumerate}
    \item  SL-WDPHN-1840-1719-K Skylink directional horn antenna, providing a gain of 19.5 dBi and a half-power angle of about 15 degrees;
    \item A-info SZ-18004000/P omnidirectional antenna with a~gain of about 5 dBi.
\end{enumerate}
We have measured the total attenuation of the used cables to be equal to 18.9 dB at the used frequency. In the specified setup, the maximal path loss that can be estimated is 120.5 dB and 149.5 dB for omnidirectional and directional antennae, respectively. These values were calculated assuming a 1\% probability of false alarm (meaning that the  noise samples are treated as wanted signals). These values are needed as censoring levels used in the estimation procedure.

The mounting points for the antennas in the vehicles were the following: A. on the rooftop, B. on the rear and the front bumper for the transmitting and receiving vehicle, respectively, C. under the chassis for measuring potential ducting below vehicles. The location of both vehicles and the accurate distance between the transmitter and the receiver have been obtained using both GPS readings and ultra-wide-band (UWB) measurements, following the rationale explained in \cite{Kryszkiewicz2020}. GPS measurements are characterized by up to a few meters innaccuracy. This becomes significant when short inter-vehicle distances are considered. The UWB measurement is much more accurate but available only for a limited range. Therefore, we have applied a combination of both methods, following \cite{Kryszkiewicz2020}, with TREK1000 devices used to process the UWB signal. 

\begin{figure}[!htb]
\centering
\includegraphics[width=0.4\textwidth]{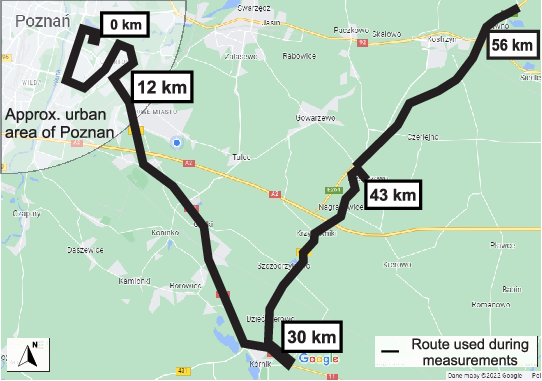}
\caption{The measurement route chosen for the drive campaigns}
\label{fig_routes}
\end{figure}

During the measurement experiment the cars forming the platoon drove along the route presented in Fig.~\ref{fig_routes}. It had begun in front of the premises of Poznan University of Technology in Poznan, Poland, with the measurements carried out across the Poznan city center and suburbs. The total length of this section was approximately 12 km. Then, after leaving the urban area, the experiment was continued along the high-speed road until the intermediate point near Kornik city had been reached. Finally, the route followed along the rural road for over 13 km. Next, the measurements were carried out on the highway section between two junctions; the platoon was driving in both directions back and forth, reaching a total travel distance of 25 km. We followed various parts of the route a couple of times, changing the antennas' mounting points and their type.  

\section{\MS{Path loss} Models}
\label{sec:plmodels}
\subsection{Prerequisities}
The ultimate goal of the conducted experiments was to identify the impact of various environmental factors on the propagation measured by means of the proposed \MS{path loss} model. First, following the approach proposed in \cite{Boban_TVT_2014} and also employed in our previous work \cite{Kryszkiewicz2022}, we tried to distinguish the line-of-sight (LOS) and non-LOS (NLOS) propagation.  To achieve this goal, the whole experiment has been recorded from the receiving car, and the video has been carefully analyzed with a millisecond precision by post-processing it frame-by-frame.
%,  and obstructed-LOS (OLOS) propagation. When it was impossible to fairly determine the channel type (e.g., due to the too-high distance), such measurements were removed from further analysis. However, it was not possible to create statistically significant 
%OLOS channel models as a result of a small number of available samples. The OLOS channel corresponds, in a real traffic scenario, to the action when one car enters or leaves the line of sight between the transmitter and receiver; such an action lasts typically very short (around 0.5-2 seconds), thus, with the measurements taken every 0.73 s, only a few samples can be obtained for one action. Thus, for the final evaluation, we split, when possible, the measurements only into two classes: LOS and NLOS, and remove all samples marked as OLOS from the analysis. 

Moreover, the samples have been precisely tagged to indicate various environmental factors and scenarios. First, all LOS and NLOS samples have been classified as "Urban" or "Rural," where the former reflects the measurements carried inside the city, and the latter - those collected outside the city. Second, we used marks: \textit{fields}, \textit{screens}, \textit{forest} and \textit{housing} to reflect the impact of the surrounding environment. In detail, \textit{fields} represent the scenario with the field areas (with no houses, bigger infrastructure, forests, or bigger bushes). \textit{Screens} illustrate the case where acoustic shields were mounted at least on one side of the street \AKK {(and the environment behind them does not matter)}. \textit{Forest} represents the case when the road went through the forestry areas, and \textit{housing} - where houses were around. A sample was tagged as described above when the receiving car entered a certain area. \AKK{Please note that some classes might be overlapping, what has to be taken into account while analyzing the results.} To minimize the impact of transition, the samples collected around the transition time have been discarded. 
%\MS{Brakuje opisu kiedy zaliczamy dany wynik do danej grupy. Np. co w sytuacji gdy nadajnik wiedzie do lasu a odbornik jeszcze nie?} \PS{No to jest chyba ten transition time?}

Please note that we have repeated our measurements, as we tried to collect as many samples as necessary to guarantee the statistical correctness of the channel modeling process. However, as the experiments have been carried out in real-life scenarios, situations occurred where the total number of samples (which could be tagged as presented above) or their distance variability was not satisfactory. Thus, we do not present PL models in such cases, marking them as unavailable.  

\subsection{Modelling Process}
Our focus is to propose a simple model that will be easily applicable in various real-life V2V scenarios where exact knowledge about the details of the environment cannot be easily acquired (to go for, e.g., ray-tracing-based solutions). Thus, from the variety of options, we selected the single-slope approach, which is widely used in different contexts. As proposed in \cite{Gustafson_ML_estimation_censoring_2015}, the path loss and shadowing were estimated using Maximal Likelihood estimators. We applied the single-slope path loss model operating in the logarithmic scale for distance \textit{d}, complemented with the log-normal shadowing component, as presented below:
\begin{equation}
    L(d) = A\log_{10}\left(\frac{d}{10}\right) + B + N(0,C),
\end{equation}
where $A\log_{10}\left(\frac{d}{10}\right)$ represents the impact of increased distance, referring to the distance of 10 m, $B$ defines the bias observed at the reference distance of 10 m, and the last component $N(0,C)$ corresponds to the impact of shadowing, modeled using a normal distribution with 0 mean and standard deviation \textit{C}. 
%Please note that the reference point $d_10$ was chosen to allow for easy comparison of pathloss at 10 m, being a typical inter-vehicle distance, between different models by looking at $B$ parameters. 
\AK{Please note that the reference distance was chosen to allow for easy comparison of path loss at 10 m between different models by looking at $B$ parameters, yet it can be easily transformed into any other needed value. The rationale behind the selection of 10 m was the following: such a distance may frequently appear in an urban scenario. Moreover, it can be treated as the minimum (possibly unsafe) distance between cars, e.g., while overtaking or in a platooning scenario.} The selected AB model offered us the necessary level of flexibility while keeping complexity at a very low level, as compared to, e.g., CI or log-distance models, where such flexibility was slightly reduced \cite{Sun2016}. Please note that the path loss depends mainly on the distance (explicitly provided in the model), and the frequency (included inside the $B$ parameter), as the measurements have been done for a single frequency, representative for the FR2 band. Should the other frequencies be measured, the frequency-dependent component could be included as well, e.g., $20log_{10}(f)$.  %\PS{A dlaczego jest reduced? Chyba trzeba wspomnieć, że tam dopasowuje się tylko nachylenie prostej.} 

\begin{figure}[!b]
\centering
% \psfrag{LoS: city (4082 samples) vs outside city (2589 samples)}[][][0.7]{ LoS: urban (4082 samples) vs rural (2589 samples)}
% \psfrag{samples: city}[][][0.4]{ \quad Samples: urban}
% \psfrag{city: PL=82.1+22.4log(d/10)+N(0,3.7)}[][][0.4]{ \quad Urban PL=82.1+22.4log(d/10)+N(0,3.7) }
% \psfrag{samples: outside city}[][][0.4]{\quad Samples: rural}
% \psfrag{outside city: PL=82.3+20.0log(d/10)+N(0,4.1)}[][][0.4]{\quad Rural PL=82.3+20.0log(d/10)+N(0,4.1)}
% \psfrag{NLoS: city (481 samples) vs outside city (1413 samples)}[][][0.7]{NLoS: urban (481 samples) vs rural (1413 samples)}
% \psfrag{samples: city}[][][0.4]{ \quad Samples: urban}
% \psfrag{city: PL=74.7+33.2log(d/10)+N(0,5.5)}[][][0.4]{ \quad Urban PL=74.7+33.2log(d/10)+N(0,5.5) }
% \psfrag{samples: outside city}[][][0.4]{\quad Samples: rural}
% \psfrag{outside city: PL=84.0+20.4log(d/10)+N(0,5.4)}[][][0.4]{\quad Rural PL=84.0+20.4log(d/10)+N(0,5.4)}
\includegraphics[width=0.5\textwidth, trim=0 0 0 0, clip]{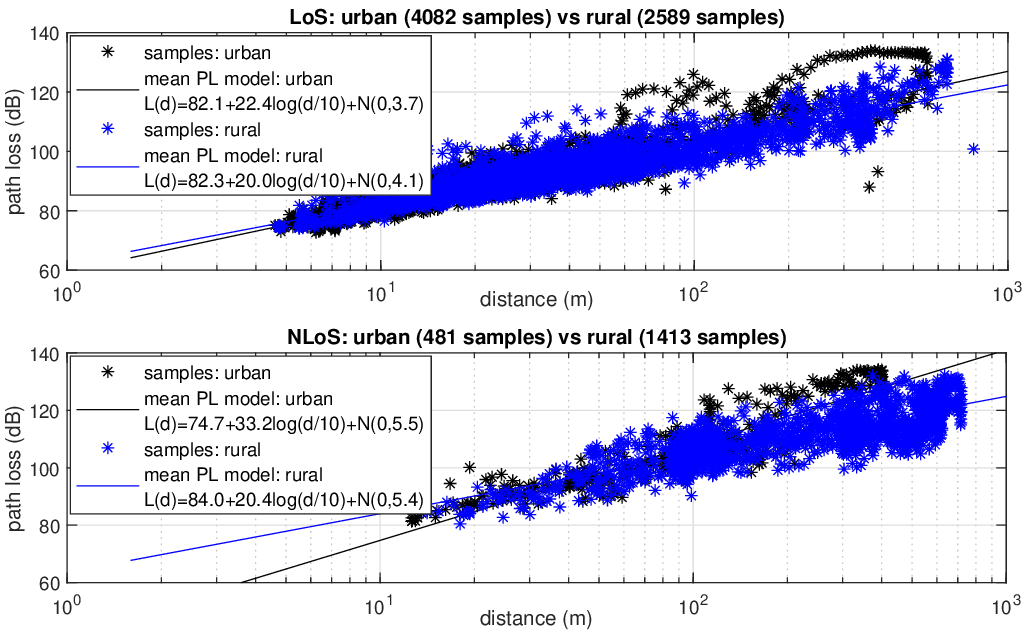}
\caption{Measurements and modeling results for omni-antenna mounted on the roof; lines in the figure do not contain the shadowing component.}
\label{fig_PL1_omni_roof_urban_rural}
\end{figure}

In Fig.~\ref{fig_PL1_omni_roof_urban_rural} an example of the ML path loss model fitting is shown. One may observe here two models (marked as \textit{urban} and \textit{rural}) for each of the two considered variants (LOS and NLOS). In the LOS case, both urban and rural PL are very similar, so the environment does not have that strong impact when there is a direct link available - we observe nearly free-space propagation, as the \textit{A} coefficient is close to 20. Next, comparing the LOS and NLOS cases, we see a more substantial shadowing impact in the latter case, which is compliant with typical models specified in, e.g., 3GPP documents \cite{3gpp38901}. Finally, in the NLOS case, we observe a higher impact of the scattering elements around vehicles in the urban case. %in consequence, higher attenuation is observed for longer distances.}
%\MS{Na wykresie LoS mamy próbki dla odleglosci okolo 2 m. Czy dla anten na dachu to nie jest troche za malo?}\PS{A distance to nie jest bumper-to-bumper? A poza tym to nie ok.2~m tylko ok. 5~m. Pozioma oś jest w skali log} 
\subsection{\AKK{Discussion on Results}}
Our measurements have considered two kinds of antennas: omnidirectional and directional. While we considered maximal antennas' gain (similarly as utilized cables attenuation) removing it from the measurements, we did not include the directivity gain of the antenna, i.e., an additional attenuation that is introduced by both TX and RX antennas not facing each other with the maximal directivity directions. The obtained path loss models include potential penalty introduced by this effect in a real traffic scenario.

Three antennas' mounting points have been investigated: on the rooftop, on the bumper, and under the chassis. The comparative results for the first two mounting locations are presented in Tab.~\ref{tab:tabmain}. Let us remember that the table consists of two parts: the left compares the PL models for urban and rural environments, whereas the right part illustrates the impact of the surrounding items in the environment (\textit{fields}, \textit{screens}, \textit{forest}, \textit{housing}). 
\begin{table*}[!t]
\caption{comparison of the path loss coefficients for various experimentation variants}
\label{tab:tabmain}
\begin{tabular}{c|lllllll|}
\cline{2-8}
& \multicolumn{7}{l|}{\cellcolor[HTML]{EFEFEF}\textbf{\begin{tabular}[c]{@{}l@{}}Path loss format $L(d) = A\log_{10}\left(\frac{d}{10}\right) + B + N(0,C)$; \quad  Presentation form \{A,B,C\}\end{tabular}}}\\
\cline{2-8}
\multirow{-2}{*}{}  & \multicolumn{1}{l|}{\cellcolor[HTML]{EFEFEF}\textbf{Channel}} & \multicolumn{1}{l|}{\cellcolor[HTML]{EFEFEF}\textbf{Urban}} & \multicolumn{1}{l||}{\cellcolor[HTML]{EFEFEF}\textbf{Rural}} & \multicolumn{1}{l|}{\cellcolor[HTML]{EFEFEF}\textbf{Fields}} & \multicolumn{1}{l|}{\cellcolor[HTML]{EFEFEF}\textbf{Screens}} & \multicolumn{1}{l|}{\cellcolor[HTML]{EFEFEF}\textbf{Forest}} & \multicolumn{1}{l|}{\cellcolor[HTML]{EFEFEF}\textbf{Housing}}\\ \hline
\multicolumn{1}{|c|}{\cellcolor[HTML]{EFEFEF}} & \multicolumn{1}{l|}{LOS} & \multicolumn{1}{l|}{\{22.4, 82.1, 3.7\}} & \multicolumn{1}{l||}{\{20.0, 82.3, 4.1\}} & \multicolumn{1}{l|}{\{19.4, 82.6, 3.9\}} & \multicolumn{1}{l|}{\{23.1, 79.9, 5.3\}} & \multicolumn{1}{l|}{\{17.6, 83.8, 3.9\}} & \multicolumn{1}{l|}{\{22.9, 82.1, 3.6\}} \\
\cline{2-8} 
\multicolumn{1}{|c|}{\multirow{-2}{2cm}{\cellcolor[HTML]{EFEFEF}\textbf{Omnidirectional at the rooftop}}} & \multicolumn{1}{l|}{NLOS} & \multicolumn{1}{l|}{\{33.2, 74.7, 5.5\}} & \multicolumn{1}{l||}{\{20.4, 84.0, 5.4\}}& \multicolumn{1}{l|}{\{20.5, 84.7, 5.0\}} & \multicolumn{1}{l|}{\{21.7, 80.4, 4.6\}} & \multicolumn{1}{l|}{\{21.2, 82.7, 6.0\}} & \multicolumn{1}{l|}{\{35.2, 72.4, 5.8\}}\\
\cline{1-8} 
%\multicolumn{1}{|c|}{\multirow{-3}{*}{\cellcolor[HTML]{EFEFEF}\textbf{Omni - roof}}}   & \multicolumn{1}{l|}{All} & \multicolumn{1}{l|}{\{A,B,C\}} & \multicolumn{1}{l||}{\{A,B,C\}} & \multicolumn{1}{l|}{\{A,B,C\}} & \multicolumn{1}{l|}{\{A,B,C\}} & \multicolumn{1}{l|}{\{A,B,C\}} & \multicolumn{1}{l|}{\{A,B,C\}}\\
\hline
\multicolumn{1}{|c|}{\cellcolor[HTML]{EFEFEF}} & \multicolumn{1}{l|}{LOS} & \multicolumn{1}{l|}{\{25.1, 79.9, 5.2\}} & \multicolumn{1}{l||}{\{19.4, 84.9, 5.9\}} & \multicolumn{1}{l|}{\{24.6, 80.6, 5.3\}} & \multicolumn{1}{l|}{\{19.6, 81.7, 3.8\}} & \multicolumn{1}{l|}{\{-,-,-\}} & \multicolumn{1}{l|}{\{29.9, 80.3, 3.8\}}\\ 
\cline{2-8} 
\multicolumn{1}{|c|}{\multirow{-2}{2cm}{\cellcolor[HTML]{EFEFEF}\textbf{Omnidirectional at the bumper}}} & \multicolumn{1}{l|}{NLOS} & \multicolumn{1}{l|}{\{14.5, 93.0, 4.1\}} & \multicolumn{1}{l||}{\{18.6, 87.7, 4.9\}} & \multicolumn{1}{l|}{\{14.5, 92.9, 4.6\}} & \multicolumn{1}{l|}{\{28.0, 76.5, 4.0\}} & \multicolumn{1}{l|}{\{-, -, -\}} & \multicolumn{1}{l|}{\{-, -, -\}}\\ 
\cline{1-8} 
%\multicolumn{1}{|c|}{\multirow{-3}{*}{\cellcolor[HTML]{EFEFEF}\textbf{Omni - bumper}}} & \multicolumn{1}{l|}{All} & \multicolumn{1}{l|}{\{A,B,C\}} & \multicolumn{1}{l||}{\{A,B,C\}} & \multicolumn{1}{l|}{\{A,B,C\}} & \multicolumn{1}{l|}{\{A,B,C\}} & \multicolumn{1}{l|}{\{A,B,C\}} & \multicolumn{1}{l|}{\{A,B,C\}}\\ 
\hline
\multicolumn{1}{|c|}{\cellcolor[HTML]{EFEFEF}} & \multicolumn{1}{l|}{LOS} & \multicolumn{1}{l|}{\{15.0, 80.6, 8.9\}} & \multicolumn{1}{l||}{\{13.9, 86.4, 7.8\}} & \multicolumn{1}{l|}{\{11.8, 86.8, 9.2\}} & \multicolumn{1}{l|}{\{15.5, 84.0, 3.7\}} & \multicolumn{1}{l|}{\{11.7, 88.6, 4.5\}} & \multicolumn{1}{l|}{\{13.8, 80.4, 9.2\}}\\ 
\cline{2-8} 
\multicolumn{1}{|c|}{\multirow{-2}{2cm}{\cellcolor[HTML]{EFEFEF}\textbf{Directional at the rooftop}}} & \multicolumn{1}{l|}{NLOS} & \multicolumn{1}{l|}{\{-, -, -\}} & \multicolumn{1}{l||}{\{16.9, 90.1, 6.4\}} & \multicolumn{1}{l|}{\{20.1, 87.4, 6.8\}} & \multicolumn{1}{l|}{\{24.4, 77.7, 4.9\}} & \multicolumn{1}{l|}{\{-, -, -\}} & \multicolumn{1}{l|}{\{-, -, -\}}\\ 
\cline{1-8} 
%\multicolumn{1}{|c|}{\multirow{-3}{*}{\cellcolor[HTML]{EFEFEF}\textbf{Dir - roof}}} & \multicolumn{1}{l|}{All} & \multicolumn{1}{l|}{\{A,B,C\}} & \multicolumn{1}{l||}{\{A,B,C\}} & \multicolumn{1}{l|}{\{A,B,C\}} & \multicolumn{1}{l|}{\{A,B,C\}} & \multicolumn{1}{l|}{\{A,B,C\}} & \multicolumn{1}{l|}{\{A,B,C\}}\\ 
\hline
\multicolumn{1}{|c|}{\cellcolor[HTML]{EFEFEF}} & \multicolumn{1}{l|}{LOS} & \multicolumn{1}{l|}{\{-, -, -\}} & \multicolumn{1}{l||}{\{5.2, 94.1, 9.1\}} & \multicolumn{1}{l|}{\{16.2, 93.5, 6.4\}} & \multicolumn{1}{l|}{\{-, -, -\}} & \multicolumn{1}{l|}{\{-, -, -\}} & \multicolumn{1}{l|}{\{-, -, -\}}\\ 
\cline{2-8} 
\multicolumn{1}{|c|}{\multirow{-2}{2cm}{\cellcolor[HTML]{EFEFEF}\textbf{Directional at the bumper}}} & \multicolumn{1}{l|}{NLOS} & \multicolumn{1}{l|}{\{-, -, -\}} & \multicolumn{1}{l||}{\{16.2, 93.5, 6.4\}} & \multicolumn{1}{l|}{\{17.3, 92.4, 6.2\}} & \multicolumn{1}{l|}{\{-, -, -\}} & \multicolumn{1}{l|}{\{-, -, -\}} & \multicolumn{1}{l|}{\{-, -, -\}}\\ 
\cline{1-8} 
%\multicolumn{1}{|c|}{\multirow{-3}{*}{\cellcolor[HTML]{EFEFEF}\textbf{Dir - bumper}}}  & \multicolumn{1}{l|}{All} & \multicolumn{1}{l|}{\{A,B,C\}} & \multicolumn{1}{l||}{\{A,B,C\}} & \multicolumn{1}{l|}{\{A,B,C\}} & \multicolumn{1}{l|}{\{A,B,C\}} & \multicolumn{1}{l|}{\{A,B,C\}} & \multicolumn{1}{l|}{\{A,B,C\}}\\
\hline
\end{tabular}
\end{table*}

Various conclusions could be drawn here by comparing the presented models. First, one may observe interesting differences in the obtained models between the mounting point for the omnidirectional antennas. As the PL models are relatively similar for the LOS case, they are quite different for the NLOS variant. For example, it is particularly visible when one compares the results for the NLOS channel in an Urban environment, where there \AKK{is almost a difference of 20 between A and B} values (${33.2, 74.7, 5.5}$ for the rooftop and ${14.5, 93.0, 4.1}$ for the bumper). It means that when the antenna is mounted on the bumper, the signal suffers from significant constant attenuation introduced by the intermediate cars (obstacle-car causing NLOS), and the impact of the distance is not that severe. It is the opposite for the rooftop mounting point, where the impact of the blocking car is much smaller, and the signal attenuates more steeply with distance. 

Next, one can compare the PL models for LOS and NLOS for rural areas when the omnidirectional antenna is mounted on the rooftop; they are very similar; thus, it could be reasonable to combine them into one model and not distinguish between the LOS and NLOS cases. Similarly, for the same mounting point of the omni-antenna and NLOS channel, the differences between the PL models achieved for non-\textit{housing} surroundings are again relatively small. 

In general, for omni-antenna mounted on a~rooftop, one can see the significant impact of the environment, especially when comparing \textit{urban} versus \textit{rural}, or \textit{housing} and \textit{screens} versus \textit{fields} and \textit{forests}. This impact is particularly noticeable in the NLOS case. With \textit{screens} or \textit{housing} as well as in the \textit{urban} scenarios, the change in attenuation of the signal with distance is much stronger ($A$ is around 20) than in \textit{rural}, \textit{fields} and \textit{forest} cases ($A$ is above 30). 

An interesting phenomenon can be observed when analyzing a variant with a directional antenna on the rooftop, and comparing the measurement results with those for an omni-antenna. First, in some cases, the shadowing component is very high (maximally even above 9 dB). 

Second, the impact of the first, distance-related component $A$, is weaker. For omni-antenna, it takes values around 20, while for directional antennas - between 11 and 16. At the same time, the bias values $B$ are comparable in both cases. 

Various conclusions can be drawn based on the achieved results; however, due to the limited space, we decided to highlight the last observation.  For the directional antenna located on the bumper, the bias coefficient $B$ is very high (above 93 dB), in both cases, where we were able to collect a reasonable number of samples (i.e. for \textit{rural} and \textit{fields}). In general, changing the location from the roof to the bumper increases $B$ coefficient.

\section{\AKK{Highlights on Specific Aspects}}
%\AKK{In this section, we discuss three specific use cases: the "below-the-truck ducting effect", the impact of blocking car(s) in motion, and the decorrelation time.}
\subsection{Antennas Mounted Below the Car}
The intention of mounting antennas below the cars was to investigate the potential of wave ducting in the tunnel between the street pavement and bottoms of the vehicles (especially trucks), as the clearance is potentially relatively ample. The following PL models have been achieved for LOS:
\begin{itemize}
\item  for urban area: $L(d) = 36.9 \cdot \log\left(\frac{d}{10}\right) + 91.7 + N(0, 3.3)$;
\item rural area $L(d) = 22.1 \cdot \log\left(\frac{d}{10}\right) + 95.3 + N(0, 3.4)$.
\end{itemize}
%wever, based on the PL measurements, there were only negligible differences between the models tailored for visual LOS and NLOS situations; thus, we propose not to distinguish between these two variants in the case of under-the-car montage, and treat it only as NLOS model due to the impact of the car chassis elements. We foresee that detailed measurements should be done to verify the impact of the size of the antenna-mounting arm.
One can notice the high values of bias $B$ (above 90 dB), and, at the same time, the high impact of distance on PL - $A$ is above 22 for rural and close to 37 for urban areas. Moreover, the shadowing variance is relatively small. Achieved results show that under-the-car communication faces very strong signal attenuation if the antennas are too close to the car chassis. 
Furthermore, we have evaluated the impact of the surrounding elements on signal propagation. We achieved reliable measurements for two variants:
\begin{itemize}
\item  for \textit{fields}  $L(d) = 23.9 \cdot \log\left(\frac{d}{10}\right) + 94.1 + N(0, 3.8)$;
\item for \textit{screens} $L(d) = 20.9 \cdot \log\left(\frac{d}{10}\right) + 94.5 + N(0, 3.7)$.
\end{itemize}

\subsection{Impact of Number of Cars on the \MS{Path Loss}}
Finally, we attempted to check if there is a significant change in PL observed when the number of blocking cars changes. In the dynamic, real-life experiment, we collected a reliable number of samples for the case where either omnidirectional or horn antennas were mounted on bumpers and when there were one (Case A) or more cars (i.e., more than one, (Case B) between the transmitter and the receiver. The results are summarized below:
\begin{itemize}
\item  Omni, Case A $L(d) = 12.5 \cdot \log\left(\frac{d}{10}\right) + 92.8 + N(0, 4.1)$;
\item Omni, Case B $L(d) = 12.0 \cdot \log\left(\frac{d}{10}\right) + 94.8 + N(0, 3.9)$;
\item  Horn, Case A $L(d) = 1.5 \cdot \log\left(\frac{d}{10}\right) + 104.3 + N(0, 5.1)$;
\item Horn, Case B $L(d) = 6.9 \cdot \log\left(\frac{d}{10}\right) + 106.5 + N(0, 5.6)$.
\end{itemize}
The results are interesting, as there is no evident difference between the cases with one or more than one blocking cars. The dominating factor is the location of the aerial. In the case of the horn antenna, one may observe the almost negligible impact of the distance ($A=1.5$ for one blocking car), whereas the bias is very high - above 100 dB. For multi-car case, the coefficient \textit{A} increases to 6.9. Let us stress that the measurements were collected for distances from 30 to 50 m for Case A and from 100 to 200 for Case B, which specifies the validity range of the proposed models.
\subsection{Comparison with 3GPP TR37.885}
We compared our results with one of the popular models defined in \cite{3gpp37885}. As the model considers various cases, here we put results for the LoS highway ($PL = 32.4 + 20\log_{10}(d_{3D}) + 20\log_{10}(f_c)$) and NLoS ($PL= 36.85 + 30\log_{10}(d_{3D}) + 18.9 \log_{10}(f_c)$) channel, for which the environment specification and the mathematical model specification (single-slope model) is similar to the one used in our paper. Fixing  $f_c=26.555$ GHz, the TR 37.885 model results for LoS case in: $A = 20.0, B=80.88, C=3$, and for NLoS case in: $A = 30.0, B=93.77, C=4$. Comparing these results with those in Tab.~\ref{tab:tabmain}, one can observe relatively high similarity between these two models. Visibly, the TR 37.885 LoS model is very close to our model for the omnidirectional antenna mounted on the rooftop: comparing the $B$ parameter, the path loss difference is lower than 1.5 dB at a 10 m distance. As in both 3GPP and our proposal $A=20$, the path loss difference will not increase for other distances. However, as different antenna mounting points impact path loss, the TR 37.885 model diverges more for the other antennas and mounting points analyzed in this paper. Our model is more specific.  
\subsection{Decorrelation Time}
Finally, we have processed the collected samples to verify what is the decorrelation time $T_d$ of the channel shadowing using Gudmundson model \cite{Gudmundson_correlation_1991}, i.e., the time when the normalized autocorrelation achieves $\frac{1}{e}$ value. Calculation details are described in \cite{Kryszkiewicz2022}. Achieved results depend on the antenna mounting point, receiver velocity $V_{RX}$ and relative velocity between the transmitter and the receiver $V_{TX-RX}$ (both velocities are quantized). The results are placed in Tab.~\ref{tab:dectime}. When the RX car does not move, $T_d$ is high (above 14 s). However, when cars are moving, $T_d$ is much smaller. This shows that even if inter-vehicle communications is blocked at some time instance by high shadowing, after a few seconds, the shadowing will be uncorrelated, possibly allowing for transmission;
%around some seconds and increases as the relative velocity increases.
higher $T_d$ is observed for the rooftop mounting point.   
\begin{table}[!h]
\centering
\caption{Decorrelation time [s] }
\label{tab:dectime}
\begin{tabular}
{|
>{\columncolor[HTML]{C0C0C0}}lllll|}
\cline{2-5}
\multicolumn{1}{c|}{\cellcolor[HTML]{FFFFFF}}& \multicolumn{1}{|l|}{\cellcolor[HTML]{C0C0C0} $V_{TX-RX} \left[\frac{m}{s}\right]$}  & \multicolumn{1}{c|}{\cellcolor[HTML]{C0C0C0}0} & \multicolumn{1}{c|}{\cellcolor[HTML]{C0C0C0}2} & \multicolumn{1}{c|}{\cellcolor[HTML]{C0C0C0}4} \\ \hline
%\multicolumn{4}{|c|}{\cellcolor[HTML]{EFEFEF}Omnidirectional rooftop} \\ \hline
\multicolumn{1}{|c|}{\cellcolor[HTML]{EFEFEF}} & \multicolumn{1}{|l|}{\cellcolor[HTML]{C0C0C0}$V_{RX} = 0 \left[\frac{m}{s}\right]$} & \multicolumn{1}{r|}{14.57}& \multicolumn{1}{c|}{--}& 22.44\\ \cline{2-5}
\multicolumn{1}{|c|}{\cellcolor[HTML]{EFEFEF}} & \multicolumn{1}{|l|}{\cellcolor[HTML]{C0C0C0}$V_{RX} = 10\left[\frac{m}{s}\right]$} & \multicolumn{1}{r|}{1.09}& \multicolumn{1}{r|}{1.00}& \multicolumn{1}{r|}{4.91}\\ \cline{2-5}
\multicolumn{1}{|c|}{\multirow{-3}{2cm}{\cellcolor[HTML]{EFEFEF}\textbf{Omnidirectional at the rooftop}}} & \multicolumn{1}{|l|}{\cellcolor[HTML]{C0C0C0}$V_{RX} = 20 \left[\frac{m}{s}\right]$} & \multicolumn{1}{r|}{1.66}& \multicolumn{1}{l|}{2.64}& \multicolumn{1}{r|}{9.04}\\ \hline
%\multicolumn{4}{|c|}{\cellcolor[HTML]{EFEFEF}Omnidirectional bumper}\\ \hline
\multicolumn{1}{|c|}{\cellcolor[HTML]{EFEFEF}} & \multicolumn{1}{|l|}{\cellcolor[HTML]{C0C0C0}$V_{RX} = 0 \left[\frac{m}{s}\right]$} & \multicolumn{1}{r|}{128.02}& \multicolumn{1}{c|}{--}& \multicolumn{1}{c|}{--}\\ \cline{2-5}
\multicolumn{1}{|c|}{\cellcolor[HTML]{EFEFEF}} & \multicolumn{1}{|l|}{\cellcolor[HTML]{C0C0C0}$V_{RX} = 10 \left[\frac{m}{s}\right]$} & \multicolumn{1}{r|}{1.41}& \multicolumn{1}{r|}{2.21}& \multicolumn{1}{r|}{5.34}\\ \cline{2-5}
\multicolumn{1}{|c|}{\multirow{-3}{2cm}{\cellcolor[HTML]{EFEFEF}\textbf{Omnidirectional at the bumper}}}  & \multicolumn{1}{|l|}{\cellcolor[HTML]{C0C0C0}$V_{RX} = 20 \left[\frac{m}{s}\right]$} & \multicolumn{1}{r|}{4.40}& \multicolumn{1}{r|}{4.37}& \multicolumn{1}{r|}{3.07}\\ \hline
\end{tabular}
\end{table}
\section{Conclusions}
\label{sec:conclusions}
In this paper, we presented the set of various single-slope PL models for V2V mmWave propagation generated based on rich measurement campaigns conducted in a real-\MS{traffic} environment. 
%Various specific conclusions can be drawn, as discussed above, but we think that the most important observations are the following. 
We identified that it is hard to identify some generic trends on how the mounting point influences the resultant observed PL. One may observe that the bias (\textit{B} component) increases in NLOS scenario when the antenna is located on the rooftop, on the bumper, and below the car. However, besides that, the overall impact is noticeable, yet the generic rules are hard to specify. Second, the potential ducting effect below the car chassis requires further analysis, as the placement of antennas on a 15 cm arm does not guarantee reliable communications. Third, there are reasonable differences in PL in urban and rural cases; however, the distinction between various surrounding elements is justified if particularly needed. Fourth, the impact of blocking cars - well reflected in static scenarios - is not that well visible in the case of antennas mounted on the bumper when cars are moving. Finally, the correlation time is around of a couple of seconds in the full mobility scenario.

\bibliographystyle{ieeetr}
\bibliography{references}

\end{document}